\documentclass[aps,prb,twocolumn,showpacs,a4]{revtex4} 

\usepackage{tabularx}

\usepackage[final]{graphicx}
\usepackage{dcolumn}
\usepackage{bm}
\def\extgra{pdf}
\def\extgra{eps}
\def\figsiz{\columnwidth}

\newcommand{\mm}[1]     {\ifmmode {#1} \else{}${#1}$\fi}
\newcommand{\mmm}[1]    {\ifmmode{}#1 \else{}${#1}$\fi}
\newcommand{\beq}[1]    {\begin{equation} \label{#1}}
\newcommand{\eeq} {\end{equation}}

\def \rA{{\mm{\langle r_A\rangle}}}

\def \LPCM{${\rm (La_{1-y}Pr_y)_{0.7}Ca_{0.3}MnO_3}$}
\def \lpcm{${\rm (La_{1-y}Pr_y)_{0.7}Ca_{0.3}MnO_3}$}

\def \lcm{${\rm La_{0.8}Ca_{0.2}MnO_3}$}
\def \lcm_ce{${\rm La_{0.5}Ca_{0.5}MnO_3}$}

\def \rA{{\mm{\langle r_A\rangle}}}

\def\Os{${\rm ^{16}O}$}
\def\Oe{${\rm ^{18}O}$}

\def\dst{ \mm{(\delta d/d)_{\rm st}} }

\begin{document}


\title{\large Evidence for strong effect of quenched correlated disorder on
phase separation and magnetism in  {$\rm (La_{1-y}Pr_{y})_{0.7}Ca_{0.3}MnO_3$}}


\author{V.~Yu.~Pomjakushin}
\affiliation{Laboratory for Neutron Scattering, ETH Zurich and Paul
Scherrer
Institut, CH-5232
Villigen PSI, Switzerland}
\author{D.~V.~Sheptyakov}
\affiliation{Laboratory for Neutron Scattering, ETH Zurich and Paul
Scherrer
Institut, CH-5232
Villigen PSI, Switzerland}
\author{E.~V.~Pomjakushina}
\affiliation{Laboratory for Developments and Methods, PSI, CH-5232
Villigen PSI, Switzerland}
\author{K.~Conder}
\affiliation{Laboratory for Developments and Methods, PSI, CH-5232
Villigen PSI, Switzerland}
\author{ A.~M.~Balagurov}
\affiliation{Frank Laboratory of Neutron Physics, JINR, 141980, Dubna,
Russia}


\date{\today}

\begin{abstract}

High resolution neutron diffraction shows that the mesoscopic
separation into ferromagnetic (FM) and antiferromagnetic (AFM) phases
and the FM transition temperature $T_C$ in the perovskite manganite
\lpcm\ strongly depend on the quenched correlated disorder. The
different disorder strengths are achieved by different procedures of
the sample synthesis and quantitatively characterized by the
micro-strain-type diffraction peak broadening. The system shifts to
predominantly one phase state with smaller $T_C$ as
the correlated disorder strength is decreased supporting the
viewpoint that the origin of phase separation in the indicated
manganite system is the correlated quenched disorder. The ground state
of an ultimately chemically homogeneous sample is FM-like containing
about 20\% of the AFM minority phase. This FM-like state can be
readily transformed to the AFM-like one having $<20$\% of FM-phase by
the decreasing of the effective charge carrier bandwidth via oxygen
isotope substitution.

\end{abstract}


\maketitle

\section{Introduction}

 The presence of the long scale percolate phase separation in the colossal magnetoresistance (CMR) manganese oxides ${\rm A_{1-x}A'_xMnO_3}$ (A is rare earth element, A$'$ is Ca, Sr, Ba) has been attracting special experimental and theoretical attention for long time. There are two main concepts of the mesoscopically inhomogeneous state in the manganites. One standpoint is that the intrinsic quenched disorder enhances the fluctuations of the competing orders near the original bi-critical point\cite{burgy01,burgy04,alonso02,blake02,akahoshi03,deteresa05, pomjakushin07}. The theoretical calculations in random field Ising model with quenched correlated disorder \cite{burgy01,burgy04} show generation of mesoscopically large clusters of both phases. In another approach, the lattice distortions and the long-range strain\cite{littlewood99, ann04, podzorov01, sharma05} are the dominant factors controlling the phase separation. The self-organized multiphase coexistence originates purely from lattice degrees of freedom and is caused by the presence of an intrinsic elastic energy landscape according to the computer simulations \cite{ann04}. There are also suggestions that the intrinsic quenched disorder and lattice distortions are interrelated in the Mn-substituted half doped manganites with low level (1-5\%) of Mn-substitution \cite{yaicle05,yaicle06,frontera08}. 

The experimental discrimination of the two models is difficult and
usually based on a concomitant effects. In this paper we give direct
evidences in favor of the disorder model by showing that the phase
separation is strongly dependent on the strength of quenched correlated
disorder, that can be quantitatively characterized by a micro-strain
diffraction peak broadening parameter.

The low bandwidth manganite family \lpcm\ (LPCM hereafter) has the
fixed optimal hole doping $x=0.3$ and variable A-cation radius \rA\
that is linearly connected with the Pr-concentration $y$. The
principal effect of decreasing \rA\ is a decrease in Mn-O-Mn bond
angle, which leads to the decrease in the electron transfer integral
between the Mn-ions. The metal-insulator boundary lays at the
Pr-concentration between $y=0.86$ and
$y=1.0$\cite{hwang95,babushkina99}. The ground magnetic state in the
metallic part of the phase diagram is an incoherent mixture of
ferromagnetic metallic FMM and antiferromagnetic insulating AFI
mesoscopically large regions.\cite{pomjakushin07} There are two
factors controlling the proportion between FMM and AFMI phases. The
first one is the effective charge carrier bandwidth $W$ that is
decreased as the Mn-O-Mn bond angle is decreased. Due to the strong
electron-phonon coupling mediated by the Jahn-Teller effect  $W$ is
also significantly decreased by the increase in the oxygen mass from
\Os\ to \Oe, thus shifting the system also towards insulating
state.\cite{pomjakushin07} The second factor that controls the phase
separation is quenched disorder, which modifies the electron transfer
integral only locally, but can give rise to the mesoscopically
inhomogeneous state in according to the model
calculations.\cite{burgy01,burgy04}

The quenched disorder in the LPCM-system is naturally present due to the dispersion of the A-cation radius, i.e. even an ideally homogeneous statistical distribution of the cations over the crystallographic A-position can result in the phase separated state. It is important to note that according to elasticity theory, any point defect will create the displacements of neighboring atoms decaying as $1/r^2$ as a function of distance $r$ creating a correlated disorder region.  In real crystals the disorder will be larger due to the presence of different kinds of crystal defects like dislocations, vacancies or composition variance. The defects are classified by their effect on the diffraction peak intensity and width\cite{krivoglaz96}: defects of the first type do not change the diffraction peak widths, but affect their intensity similar as atomic displacement parameter (ADP) does and shift the peak positions. The second type defects give rise to the peak broadening. Any finite defect is the first type defect and will effect only on the integrated peak intensities. To get the line broadening from the finite defects in 3D crystal the displacements of neighboring atoms has to decrease at lower rate than $r^{-3/2}$. Any solid solution independently on the partial atomic concentrations will give narrow diffraction peaks.\cite{krivoglaz96} Thus, in case of an  ideal homogeneous statistical distribution of the A-cations one expects to have an increase in the ADP and narrow resolution limited Bragg peak widths. Large scale defect, such as dislocation, is the second type defect and will create the broadening of the diffraction lines. 
Possibly some clusterizations of particular A-cations can act as the second type defects. In a powder diffraction experiment additional broadening can occur also for the first type defects  due to the fluctuation of the defect concentration. For the homogeneous solid solutions this broadening is negligibly small but will have noticeable value in the presence of new phase particles.\cite{krivoglaz96}  In our case the clusterization of particular A-cations can play the role of the "new phase particle". We have prepared three types of samples with different correlated quenched disorder strength that was quantitatively characterized by micro-strain type line broadening and have studied the phase separation effect and magnetism as a function of the disorder.

\section{Samples. Experimental}
\label{exp}

The samples of \lpcm\ with y=0.8 have been synthesized by a solid
state reaction using $\rm La_{2}O_{3}$,  $\rm Pr_{6}O_{11}$, $\rm
MnO_{2}$ and $\rm CaCO_{3}$. The samples, which are denoted as
o-samples, were calcined at temperatures 1000-1300$^{\rm o}$C
for 100h with 3 intermediate grindings. The final \Oe- and
\Os-samples were obtained via respective oxygen isotope exchange
in closed quartz tubes in parallel under the controlled gas
pressure slightly above 1 bar at 1000$^{o}$C during 40h. Second
type is t-sample that has been additionally sintered at higher temperature 
1500$^{\rm o}$C during 40h. To further increase chemical
homogeneity we applied a procedure similar to the one described in
Ref.~\onlinecite{collado} to synthesize the samples that we call
m-samples. Stoichiometric amounts of the
starting materials (15 g) were mixed in agate mortar and
calcinated at 950$^{o}$C for 12 h (as for standard o-samples).
At the next step preliminarily powderized sample was placed in
Fritsch planetary mill (Pulverisette 5, agate balls and grinding
bowl) and 15 ml isopropanol was added as homogenizing liquid. The
sample was milled totally 4 h at 100 rpm with changing the
spinning direction every hour. The obtained suspension was dried
at 100 $^{o}$C and then annealed at 950$^{o}$C for 20 h. 
The sample was milled in the Fritsch mill at the same conditions as 
at the previous step, dried, pressed into the pellets and 
annealed at 1450$^{o}$C for 15 h .
Finally the pellets were grounded in agate mortar, pelletized and
sintered 1500 $^{o}$C for 12 h.

The chemical homogeneity or composition variance in the solid state
synthesis method is expected to be better if the Fritsch planetary
mill is used because the minimal comminuted particle sizes are 0.1$\mu$m that is
significantly smaller than 5-10$\mu$m if grounded in a standard agate
mortar. The higher sintering temperature is also expected to be a
factor improving the homogeneity due to higher diffusion coefficient.

The \Oe\ -samples had 80\% of \Oe-isotope, measured by the weight
gain after the oxygen exchange. The control weighing of the
\Os-sample gave the same mass within accuracy $0.03$\%. The mass
of each sample was about 2~g. The oxygen content in all the
samples was determined by the thermogravimetric hydrogen
reduction\cite{conder02} and amounted to 3.003(5). The $ac$
magnetic susceptibility $\chi(T)=\chi'(T)+i\chi''(T)$ was
measured in zero external field with amplitude of the $ac$ field
10~Oe and frequency 1~kHz using Quantum Design PPMS station.
Neutron powder diffraction experiments were carried out at the
SINQ spallation source of Paul Scherrer Institute (Switzerland)
using the high-resolution diffractometer for thermal neutrons
HRPT\cite{hrpt} ($\lambda=1.866, 1.494$~\AA, high intensity mode
$\Delta d/d\geq1.8\cdot10^{-3}$), and the DMC diffractometer
\cite{dmc} situated at a supermirror coated guide for cold
neutrons at SINQ ($\lambda=2.56$~\AA). All the temperature scans
were carried out on heating.  The refinements of the crystal and
magnetic structure parameters were carried out with {\tt
FULLPROF}~\cite{Fullprof} program, with the use of its internal
tables for scattering lengths and magnetic form factors.

\section{Results and discussion}
\label{res}

\begin{figure}[t]
  \begin{center}
    \includegraphics[width=\figsiz]{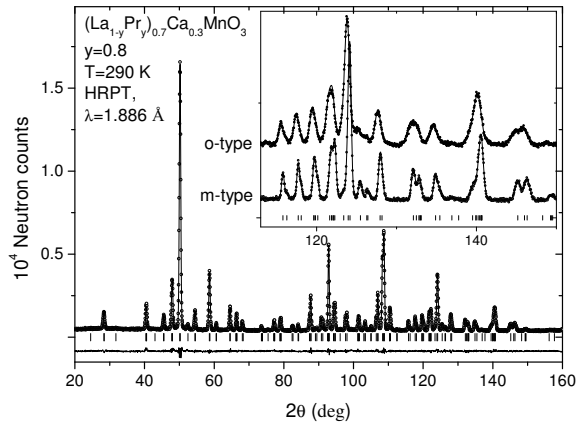}
  \end{center}

 \caption{
An example of the Rietveld refinement pattern and
difference plot of the neutron diffraction data for the \lpcm\ m-type sample with the Pr-content $y=0.8$. The inset shows the fragments of diffraction patterns for both o- and m-type samples illustrating much narrower peak widths in the m-sample.}
  \label{riet_ex}
\end{figure}

\begin{table}[b]
\label{tab1}
\caption{
The microstrain line broadening parameter \dst\ determined from
NPD at room temperature  in \LPCM\  samples. The number in the
sample name stands for $y\cdot100$ and the letter denotes the sample
type. All the samples are with natural oxygen isotope \Os\ except
the 80-m18 that is \Oe-substituted. The data for the samples 80-o and 20-o are from 
Ref.~\onlinecite{pomjakushin07}. The
Curie transition temperatures $T_C$ and r.m.s. variance of the
Gaussian distribution of the transition temperatures $\delta T_C$
were determined from the fits of neutron diffraction intensities
$I(T)$.\cite{pomjakushin07} $m_F$ and $m_A$ are effective ferro-
and antiferromagnetic moments at $T\leq15$~K determined by NPD.
}

\begin{center}
\begin{tabularx}{\linewidth}{XXXXXX}

\hline\hline
Sample &  $10^4\cdot\delta d/d$ & $T_{C}$, K & $\delta T_{C}$, K &
$m_{A},\mu_B$  & $m_{F},\mu_B$ \\ \hline

80-o   &  26(2) & 120(2) & 23(2) &  1.65(2)  & 3.03(2) \\ 
80-t   & 22(3) & 106(1)& 11(2) &  1.01(5)  & 3.21(4) \\ 
80-m   & 7.0(1.8) &90.4(4) & 3.2(4)&  0.93(3)  & 3.40(2) \\ 
80-m18 & 8.8(1.4) &85(2)   & 3.3(4)& 2.05(2)  & 1.47(3)  \\ 
20-o   &  26(2) &240.3(1)&  0    &  0.15(2)  & 3.57(2) \\ 
20-t   &  12(3)   & 223(1) &  0    &  -        & 3.49(5) \\
\hline\hline
\end{tabularx}

\end{center}
\end{table}

The crystal structure at all temperatures is well refined in single phase in space group $Pnma$  with the standard for these compounds model.\cite{Bala01LPCM_crystal} An example of the Rietveld refinement pattern and difference plot of the neutron diffraction is shown in Fig.~\ref{riet_ex}. The structure disorder can be characterized by atomic displacement parameters (ADP) and microstrain type line broadening. We indeed observe large ADP $B=0.90(2), 0.50(2), 1.00(3)$~\AA$^2$ for A-cation, Mn and oxygen atoms, respectively in \lpcm\ (y=0.8). The $B$-values are more than two times larger than in a "defect free" isostructural $\rm LaMnO_3$ that has $B=0.34(2), 0.21(3), 0.47(3)$~\AA$^2$, respectively.\cite{carvajal98} Both $\rm LaMnO_3$ and LPCM compounds have the same Jahn-Teller-type (JT) transition at high temperature (~800K) from pseudo-cubic to the antiferrodistortive orbital ordered state with completely filled $z^2$-type orbital at room temperature \cite{pomjakushin07}. The enhanced ADP are naturally expected due to the distribution of the cations with significantly different atomic radii over the A-positions and the presence of 30\% of non-JT active manganese ions $\rm Mn^{4+}$ that also act as the defects. In all three types of samples with y=0.8 the ADP values are the same within experimental errorbars implying that the point-like quenched disorder does not depend on the sample type. The parameter characterizing the correlated quenched disorder is the microstrain-type line broadening parameter \dst\ that has been determined from the refinement of the diffraction patterns as described in Ref.~\onlinecite{pomjakushin07}.  The refined values of \dst\ are shown in Table~1. For all samples, the apparent crystalline sizes $L$ are refined to the values larger than $5\times10^3$~\AA\ with the errorbars of the same order of magnitude giving only a small contribution to the line broadining (assuming $1/L=0$ increases the refined micro-strain values by less than 10\%). One can see that the additional thermal treatment (t-samples) slightly decreases the micro-strain, whereas the use of the planetary mill for the m-samples drastically decreases \dst\ down to $7\times10^{-4}$ that is illustrated by the inset of Fig.~\ref{riet_ex}. M-samples have probably an ultimate chemical homogeneity because the microstrain is close to the one expected from the twin tension due to the high temperature pseudocubic-orthorhombic structure transition.\cite{pomjakushin07} T- and o-samples can have some clusterization of the La/Ca/Pr atoms because in the solid state synthesis the high temperature reaction takes place between small solid pieces of the comminuted materials.

\begin{figure}[h]
  \begin{center}
    \includegraphics[width=\figsiz]{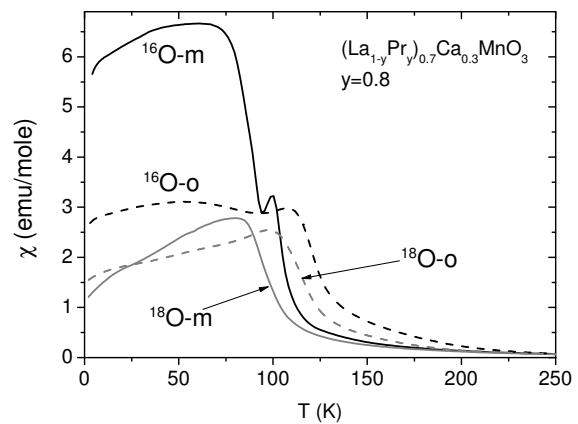}
  \end{center}

 \caption{Real $\chi'$ part of
the $ac$ magnetic susceptibilities are shown as a function of
temperature for \lpcm\ samples with the Pr-content $y=0.8$. 
M- and o-samples are shown by solid and dashed lines
respectively. \Oe-substituted samples are shown by gray lines.
}
  \label{x80all}
\end{figure}

The ac-magnetic susceptibility as a function of temperature $\chi(T)$ for both o- and m-samples is shown in Fig.~\ref{x80all}. One can see that $\chi(T)$  dependence in the milled m-sample possesses quite sharp feature around 100~K and significantly larger absolute values of $\chi^{'}$. The reason for the different behaviors of m- and o-samples is the different proportion between FM- and AFM-fraction as we show below. An important consequence of this ``sample effect'' for the manganite physics is that the temperature dependences of various macroscopic quantities such as magnetic susceptibility or electrical resistivity will not be reproducible in different experimental works if the quenched correlated disorder is not controlled. Hence, different experimental data cannot be easily compared with each other and compared with the theory predictions. Note that all structure parameters of o- and m-samples (e.g. bond lengths and angles) are the same within the errorbars. A similar observation was reported for the charge ordered $\rm Nd_{1/2}Sr_{1/2}MnO_3$, where the microstrain and phase separation onto FM and AFM phases below 150~K depended on the sample preparation procedure.\cite{woodward99}

\begin{figure}[t]
  \begin{center}
    \includegraphics[width=\figsiz]{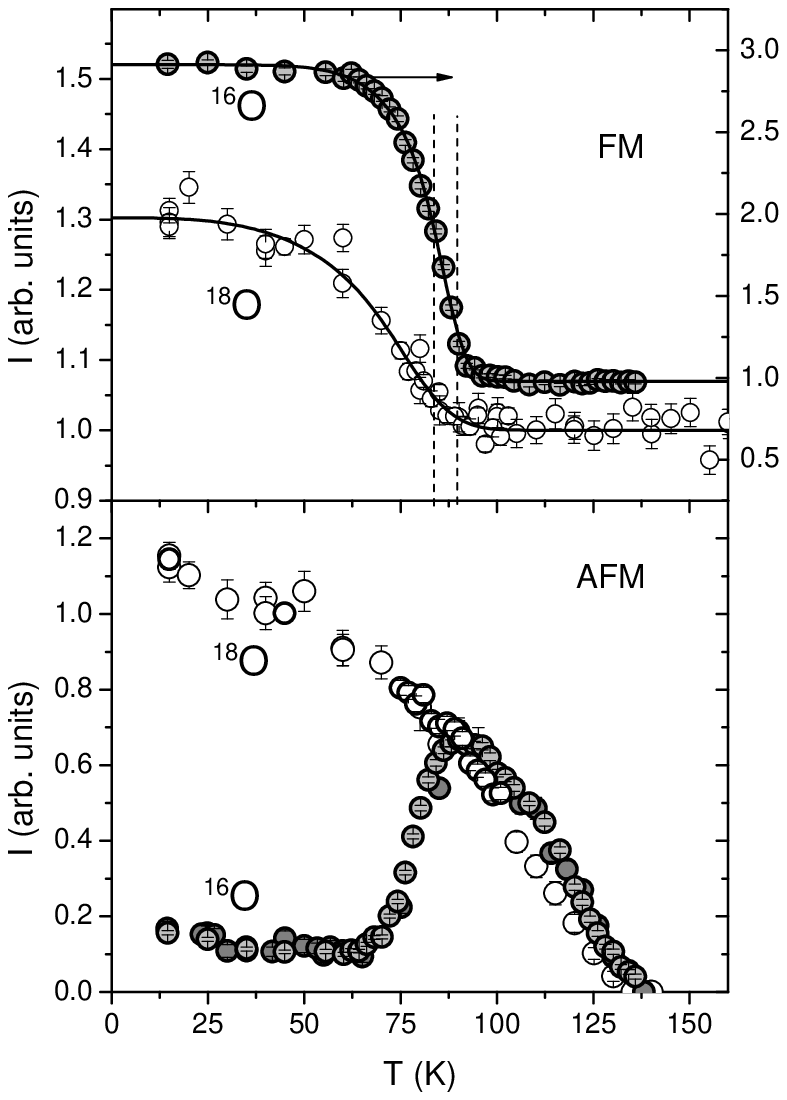} 
  \end{center}

  \caption{ 
Temperature dependences of the integrated intensities of the
selected ferro- (FM) and antiferromagnetic (AFM) diffraction
peaks in the \lpcm\ m-samples $y=0.8$. The data were collected on
heating. \Os\ and \Oe-samples are shown by closed and open
symbols respectively.  The lines are fits to the formula
described in the Ref.~\onlinecite{pomjakushin07}. Note that the
FM-intensity of the \Oe-sample is significantly smaller than the
\Os-one and has different y-axis. The vertical lines indicate the
refined positions of the FM transition temperatures.
}
  \label{giall}
\end{figure}

The magnetic phase separated state consists of the AFM pseudo-CE phase
with Mn-spin directed along $b$ and FM phase with Mn-spin in
$(ac)$-plane similar to the one reported in
Ref.~\onlinecite{Bala01LPCM_magnetic}. The FM transition temperatures $T_C$
(Table~1) were determined from the magnetic Bragg peak intensities
$I(T)$ (Fig.~\ref{giall}) by the fit procedure described in
Ref.~\onlinecite{pomjakushin07}. One can see that the transition
temperatures $T_C$ and the transition width $\delta T_C$ are decreased
as the microstrain \dst\ is decreased. The counterintuitive decrease
in $T_C$ however finds its explanation in the 3d-correlated disorder
double exchange model.\cite{bouzerar07} The calculations show that at
hole densities lower than $x=0.5$ where the couplings are strongly
inhomogeneous, $T_C$ in the inhomogeneous system is larger than
in the homogeneous one.\cite{bouzerar07} In case of uncorrelated Anderson disorder the
effect would be opposite, i.e. the inhomogeneous system would have
lower transition temperature. To further prove this
effect we have additionally prepared and studied \lpcm\ (y=0.2)
t-sample that is located in the ferromagnetic metallic part of the
phase diagram.\cite{pomjakushin07} We have found that it shows
essentially the same behavior (Table 1) - after the additional
annealing of the t-sample the transition temperature $T_C$ is
decreased together with the microstrain \dst.

\begin{figure}[t]
  \begin{center}
    \includegraphics[width=\figsiz]{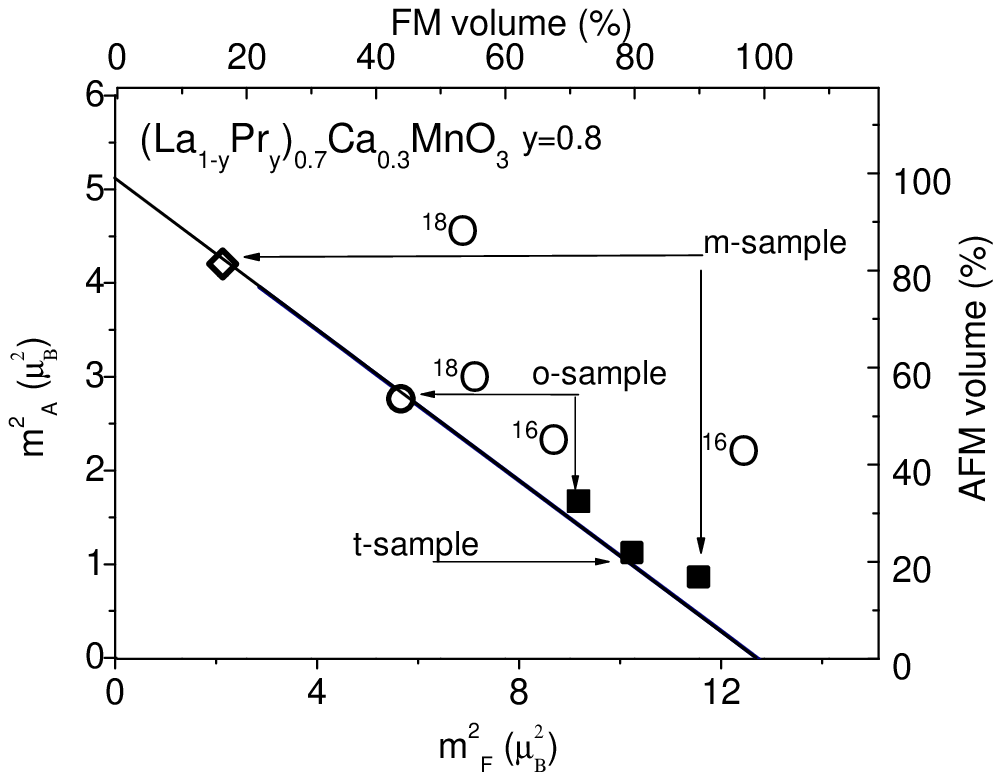} 
  \end{center}

  \caption{Effective antiferromagnetic (AFM)
  $m_{A}^2$ as a function of the effective ferromagnetic (FM)
  moment $m_{F}^2$ for the \lpcm\ samples with $y=0.8$. The
  top-$x$ and right-$y$ axes show respective FM- and AFM-phase
  fractions. The closed and open symbols represent the \Os\ and
  \Oe\ samples, respectively. Three types of samples: o-, t- and
  m-sample with different strength of correlated quenched disorder are indicated by
  arrows. The data for o-samples and the straight line, which is a fit result are 
  taken from Ref.~\onlinecite{pomjakushin07}.} 

\label{maff}
  \end{figure}

Figure \ref{maff} shows the fractions of the FM- and AFM-phases in the
samples of \lpcm\ $y=0.8$ calculated from the effective moments from
Table 1. The straight line is taken from
Ref.~\onlinecite{pomjakushin07} and represents a fit to the
experimental effective moments $m_A^2(m_F^2)$ for the whole series of
o-samples with $y=0.2-1$. One can see that the \Os-samples gradually
move along this straight line towards the FM-state as the strength of
the disorder is decreased (from o- to t- and to m-sample). This shows
that the stronger the correlated disorder the larger the separation
into FM and AFM phases. An additional evidence comes from the effect
of oxygen isotope substitution also shown in the Fig.~\ref{maff}. A
decrease in the effective bandwidth in the \Oe-samples shifts the
phase balance toward the AFM-state, as expected. But in addition, the
low defect \Oe\ m-sample displays much larger AF-fraction in
comparison with the o-samples, i.e. again the sample with minimal
correlated disorder tends to be in single phase state.  However, even
the ultimately chemically homogeneous m-samples still contain 10-20\%
of the minority phase justifying that the mesoscopically inhomogeneous
magnetic state is an intrinsic property of \lpcm.

One could suggest that the single crystals might be the most suitable candidates with the best homogeneity to verify further the effects of correlated quenched disorder. Due to the high melting temperatures of LPCM a reasonably large crystal can be grown only by traveling solvent floating zone technique (TSFZ).  
A single crystal of \lpcm\ (y=0.7) grown by TSFZ has been studied  by magneto-optical imaging technique \cite{tokunaga04}. Figure~4 of the above reference nicely shows the presence of mesoscopic phase separation onto FM and AFM phases at T=34~K. From this figure, one could estimate that the fraction of the AFM phase is more than 20\%. Thus the phase separation is even more pronounced than in the ultimately homogenous powder m-samples implying that the defect concentration in the crystal is not smaller. It is known that the crystals grown by the TSFZ or similar technique from melt can be inhomogeneous and contain many defects mainly due to large temperature gradients \cite{West}. The polycrystalline materials that are sintered at high temperatures and slowly cooled down might have less defects than the single crystals grown by TSFZ, provided that the pulverized particle sizes are small enough.

\section{Conclusions}
 
We have studied the effect of correlated quenched disorder on the
phase separation in perovskite manganese oxide \lpcm. The different
disorder strengths were achieved by different procedures of the sample
synthesis and quantitatively characterized by the micro-strain-type
diffraction peak broadening. We have found that the decreasing of the
correlated disorder pushes the system to predominantly one phase and
decreases ferromagnetic Curie temperature in accordance with the
theory predictions.\cite{burgy04,bouzerar07} The ground state of an
ultimately chemically homogeneous sample (i.e. with statistical
distribution of La/Pr/Ca cations over A-position) is still phase
separated one but with the dominant magnetic phase occupying
$80\!-\!90$\% of the volume. The dominant phase is the ferromagnetic one
for the sample with natural oxygen, but it can be readily transformed to
the antiferromagnetic phase by the decreasing of the effective charge
carrier bandwidth via \Os/\Oe-oxygen isotope substitution.

An important consequence of the effect of the correlated disorder for
the manganite physics is that the temperature dependences of various
macroscopic quantities such as magnetic susceptibility or electrical
resistivity will not be reproducible in different experimental works
and cannot be correctly compared with the theory predictions if the
correlated quenched disorder strength is not controlled.

\section*{A{\lowercase{cknowledgements}}}

This study was performed at Swiss neutron spallation source SINQ of Paul
Scherrer Institute PSI (Villigen, PSI). Financial support by the NCCR
MaNEP project is gratefully acknowledged.

\bibliography{../../../refs/refs_general,../../../refs/refs_manganites,../../../publication_list/publist2007,../../../publication_list/publist2006,../../../publication_list/publist2005,../../../publication_list/publist}

\end{document}